\documentclass[10pt]{iopart}

\usepackage{xcolor}
\usepackage{graphicx}
\begin{document}

\title[]{Nature of excitons in PPDT2FBT: PCBM solar cell: Role played by PCBM}

\author{Subhamoy Sahoo$^{1}$, Dhruvajyoti Barah$^{2}$, Dinesh Kumar S$^{1}$, Nithin Xavier$^{2}$, Soumya Dutta$^{2}$, Debdutta Ray$^{2}$, Jayeeta Bhattacharyya$^{1*}$}

\address{$^{1}$Department of Physics, IIT Madras, Chennai 600036, India}
\address{$^{2}$Department of Electrical Engineering, IIT Madras, Chennai 600036, India}

\ead{jayeeta@iitm.ac.in}
\vspace{10pt}
\begin{indented}
\item[]June 2022
\end{indented}

\begin{abstract}
In organic semiconductor based bulk heterojunction solar cells, the presence of acceptor increases the formation of charge transfer (CT) excitons, thereby leading to higher exciton dissociation probabilities. In this work we used steady state EA measurements to probe the change in  the nature of excitons as the blend composition of the solar cell active layer material is varied. We investigated blends of poly[(2,5-bis(2-hexyldecyloxy)phenylene)-alt-(5,6-difluoro-4,7-di(thiophen-2-yl)benzo[c]-[1,2,5]thiadiazole)] (PPDT2FBT) and (6,6)-Phenyl C71 butyric acid methyl ester (PCBM). Analysis of the EA  spectra showed that in presence of fullerene based acceptor, like PCBM, CT characteristics of the excitons were modified, though, no new CT signature was observed in the blend. Enhancement in the CT characteristic in the blend was reflected in the photoluminescence (PL) measurements of the blends, where, PL quenching of $\sim$ 63\% was observed for 1\% PCBM. The quenching reaches saturation at about 20\% PCBM. However, the maximum efficiency of the devices was obtained for the blend having 50\% PCBM. Comparing experimental results with simulations, the variation of the device efficiency with PCBM percentage was shown to be arising from multiple factors like increase in polarizability and dipole moment of excitons, and the efficiency of the carrier collection from the bulk of the active layer.
\end{abstract}

%
%
%
%
\ioptwocol

\section{\label{sec:level1}Introduction}
When organic semiconductors are excited optically, typically, Frenkel and charge-transfer excitons are formed\cite{pope1999electronic}. The generated excitons can either dissociate into free electron-hole pairs, radiatively recombine to generate photons, or de-excite non-radiatively via phonons. For organic solar cells, it is desirable to minimize the radiative and non-radiative recombination processes to increase the probability of the generation of free electron-hole pairs \cite{babusenan2019investigation}, which contribute to the photocurrent. Donor-acceptor heterojunctions assist in the exciton dissociation \cite{lai2013properties, devizis2015dissociation} and are, therefore, widely used for fabricating organic solar cells. The efficiency of the solar cell depends mainly on three processes - (i) absorption of photons and formation of excitons, (ii) dissociation of excitons and generation of free carriers, and (iii) collection of the carriers at the electrodes \cite{bassler2015hot, wang2019ultrafast, devizis2015dissociation}. Upon photoexcitation, the excitons generated in the donor molecule can transfer its electron to the nearest acceptor, thereby creating charge-transfer excitons. The lower energy of the acceptor level compared to the donor facilitate the process. The generated excess energy helps to break the CT excitons \cite{dimitrov2012energetic}. Also, the change in polarizability in the system in the presence of acceptor helps in exciton dissociation \cite{guan2018evidence}. Recent studies show that in organic blends, the free carriers are generated directly by ultrafast processes \cite{zhong2015ultrafast}. The created free electrons and holes are then collected at the electrodes. The efficiencies of all these mechanisms depend on the composition, thickness, and morphology of the films \cite{gao2018production, chen2020ultrafast}. 

The formation of CT excitons in organic blends is one of the important processes which determines the performance of the organic photovoltaic devices. The signature of charge transfer states in organic blends can be identified using various methods such as theoretical calculation using density functional theory \cite{izquierdo2019theoretical}, photoluminescence spectroscopy \cite{liu2016charge}, electroabsorption (EA) spectroscopy. The EA spectroscopy is widely used to probe excitons in pristine organic semiconductors and their blends. In EA spectroscopy, the change in the absorption coefficient of a material is measured in the presence of an externally applied electric field. Being a modulated measurement, the lineshape of the EA spectrum resembles the derivatives of the zero field absorption spectrum. From the spectrum analysis, the type of excitons generated in the system can be identified \cite{sahoo2022investigation}. 

PPDT2FBT, which is used as active material in this study, is an ideal candidate for the organic solar cell due to its high absorption coefficients over the entire visible spectrum \cite{nguyen2014semi}. It has high hole mobility and excellent thermal stability \cite{nguyen2014semi,ko2016photocurrent,koh2017enhanced}. PCBM, which has high absorption coefficient in the UV-blue region, is widely used as an acceptor material in solar cells. In this study, we show how PCBM influences the excitonic nature in PPDT2FBT: PCBM blends and affects the performance of the device. Electroabsorption spectroscopy (EA) was used to examine the characteristics of the generated excitons in different blend compositions. The performance of the devices of different blend compositions were estimated from photocurrent  measurements. PCBM was found to play multiple roles in improving the efficiency of the devices.

\section{\label{sec:level2}Experimental Details}
\subsection{Sample Fabrication}
Poly[(2,5-bis(2-hexyldecyloxy)phenylene)-alt-(5,6-difluoro-4,7-di(thiophen-2-yl)benzo[c]-[1,2,5]thiadiazole)] (PPDT2FBT) (Brilliant Matters) and (6,6)-Phenyl C71 butyric acid methyl ester (PC$_{71}$BM) (Lumtec) and their blends were used as active layers in this study. We use PCBM to represent PC$_{71}$BM. The blends were prepared by mixing pristine PPDT2FBT and PCBM by weight percentage as required. Dichlorobenzene was used as the solvent for preparing the solution of pristine and blends. The concentration of the prepared solutions was 15 mg/ml. The films were prepared by spin coating the solution ($\sim$ 60 $\mu$L) on the pre-cleaned substrates at 1000 rpm followed by annealing at 130 $^{0}$C for 10 minutes. 

For absorption and PL measurement, the film was prepared on a cleaned glass substrate. For EA measurement, the film was fabricated on PEDOT: PSS layer (30 nm), coated on a cleaned patterned ITO coated glass. Then, a semitransparent Al pad ($\sim$ 11 nm thick and 2 mm width) was deposited on top of it by thermal evaporation at a base pressure of $\sim$  $5 \times 10^{-7}$  mbar. For photocurrent and PCE measurement, a thick Al ($\sim$ 80 nm) was used instead of semitransparent Al, and a thin layer of LiF (1 nm) was deposited between the active layer and Al electrode. As the photocurrent and power conversion efficiency (PCE) measurements were performed at ambiance, the devices were encapsulated with cover glass and epoxy resin to prevent degradation. 

We prepared films of 9 different blends (PPDT2FBT: PCBM) having PCBM percentage (weight) 1, 2, 3, 5, 10, 20, 33, 50 and 66 respectively apart from pristine PPDT2FBT and PCBM.

\subsection{Sample Characterization}
The PL of the samples was measured using Horiba Jobin Yvon spectrometer (HR800) in backscattered geometry for 488 nm excitation. The absorption spectrum was measured using a homemade absorption spectrometer. A Xe lamp (75 W) was used as the broadband source. The quasi-monochromatic light was shone on the sample selected by the monochromator (McPherson 207). A Si photodiode measured the intensity of the transmitted light, was connected to a lock-in amplifier (SR-830) through a preamplifier (SR-570). The incident light on the sample was chopped using a mechanical chopper at 81 Hz, which was fed as the reference frequency of the lock-in amplifier. The EA was measured in transmission mode using the same configuration as the absorption measurement. In this case, the device replaced the thin film and was kept inside a vacuum chamber at $10^{-2}$ mbar pressure. The light was shone on the device from the ITO side. The bias (superposition of AC and DC voltage) was applied externally using a function generator. A reference signal, having a frequency the same as the applied AC bias, was fed to a lock-in amplifier to measure the change in transmittance in the presence of the field ($\Delta T = T_{F} - T$). The background ($T$) was measured similarly as the transmittance of the thin films. 

The PCE of the devices was measured under 1 sun illumination (AM1.5) using PET CT50AAA solar simulator at 0 V bias. The integrated photocurrent was measured by Keithley 2400 source meter unit (SMU). 

The photocurrent response spectra of the devices were measured using Bentham PVE300 photovoltaic characterization system. A pulsed monochromatic light was shone on the device and the generated photocurrent was measured using lock in amplifier. The bias to the devices was applied using Keithley 2400 SMU. The intensity of the incident light was measured using a calibrated Si photodiode. All the measurements were done in reverse bias.

\section{\label{sec:level3}Results and Discussion}

\begin{figure}
	\centering
	\includegraphics[width=\linewidth]{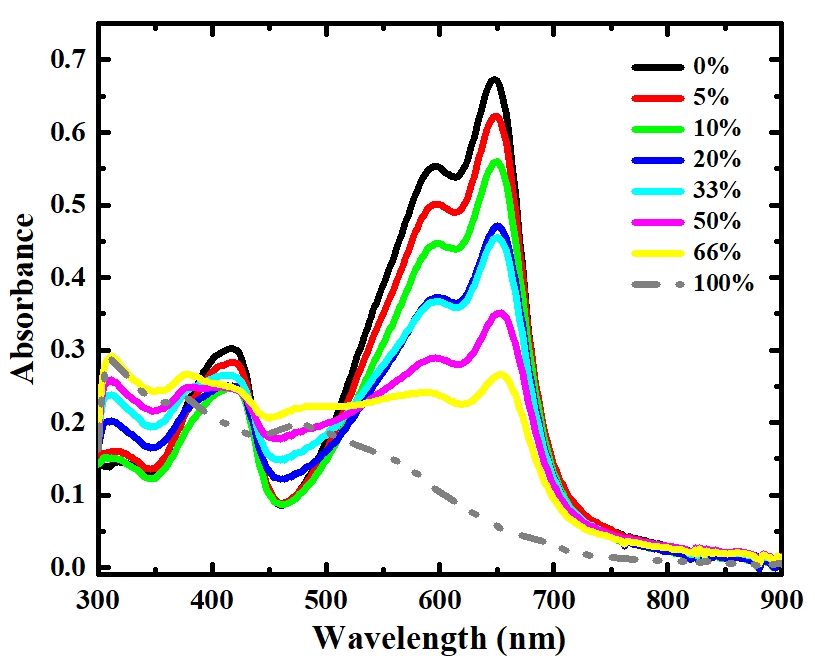}
	\caption{ The absorption spectra of PPDT2FBT (solid black line), PCBM (dash-dot gray line) and their blends. The legends show the weight \% of PCBM in the blend.  }
	\label{Fig:PPDT2FBT_Blend_Absorption}
\end{figure}

The absorption spectra of the pristine PPDT2FBT, PCBM thin films, and their blends are shown in Figure \ref{Fig:PPDT2FBT_Blend_Absorption}. The absorbance of the blends changes with the amount of PCBM. The absorption of the blend at the lower energy primarily arises from PPDT2FBT, whereas the absorption at high energy comes from both PPDT2FBT and PCBM. The optical properties of the blends changes due to the presence of PCBM, which acts as an acceptor.

\begin{figure}[h!]
	\centering
	\includegraphics[width=\linewidth]{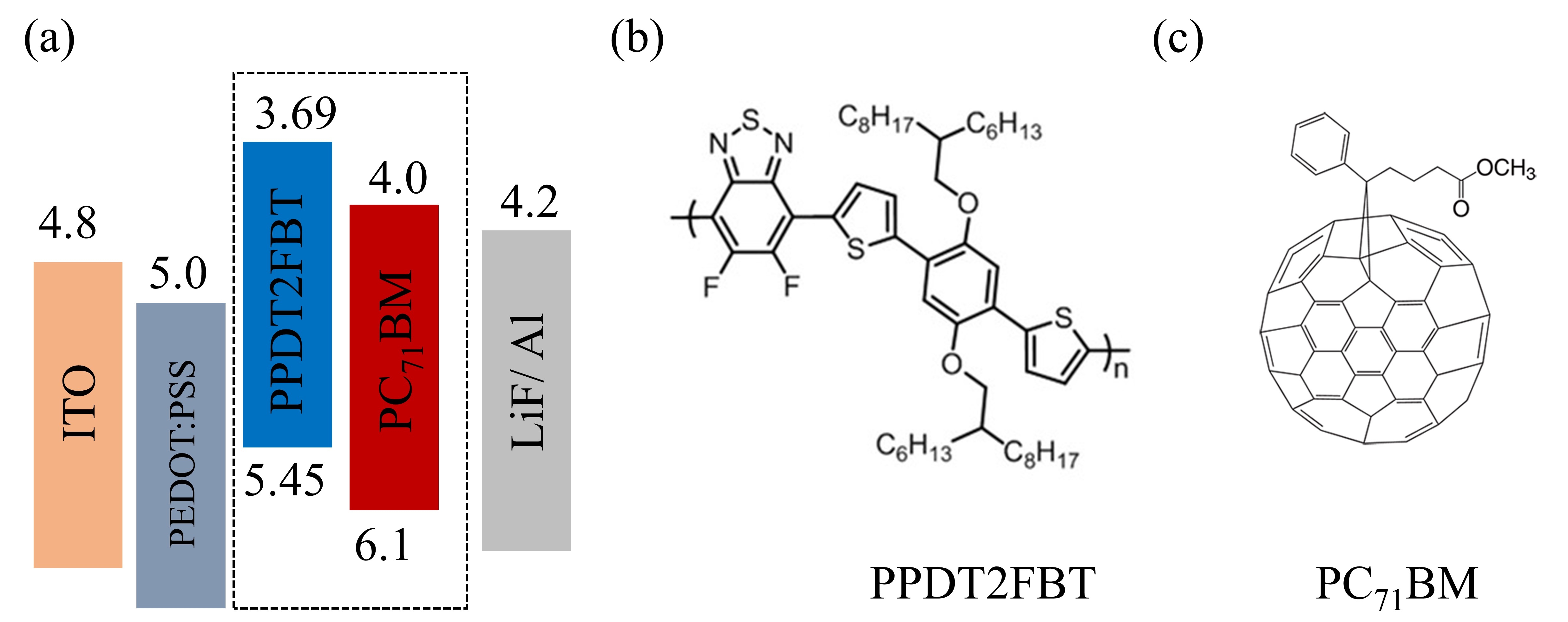}
	\caption{(a) The energy-band diagram of the materials used in devices, the chemical structure of (b) PPDT2FBT and (c) PC$_{71}$BM  }
	\label{Fig:PPDT2FBT_Blend_Energy_diagram}
\end{figure}

PPDT2FBT and PCBM form a type-II heterojunction as shown in the energy band diagram (Figure \ref{Fig:PPDT2FBT_Blend_Energy_diagram}). Therefore, the presence of acceptor in the blend helps in formation of interfacial charge transfer excitons \cite{lai2013properties}. In our earlier studies \cite{sahoo2022investigation}, we showed that, for pristine PPDT2FBT, the generated excitons had mixed nature i.e. both Frenkel and charge transfer (CT) characteristics. In blends, the nature of photogenerated excitons is expected to be modified due to charge transfer between PPDT2FBT and PCBM. Steady state electroabsorption (EA) spectroscopy can be used to study the nature of excitons in these semiconductor films \cite{guan2018evidence, wan2021direct}. In the case of Frenkel excitons, the change in polarizability ($\Delta P$) between the ground state and excited state energy level is significant. Whereas the change in the dipole moment ($\Delta \mu$) indicates the formation of CT excitons in a transition \cite{guan2018evidence, sahoo2022investigation}. Both these contributions have distinct lineshape, revealing the nature of the exciton related to any particular transition. The line shapes of the EA spectra resemble the first and second derivatives of the absorption spectrum for Frenkel and CT excitons, respectively.

\begin{figure}[h!]
	\centering
	\includegraphics[width =0.9 \linewidth]{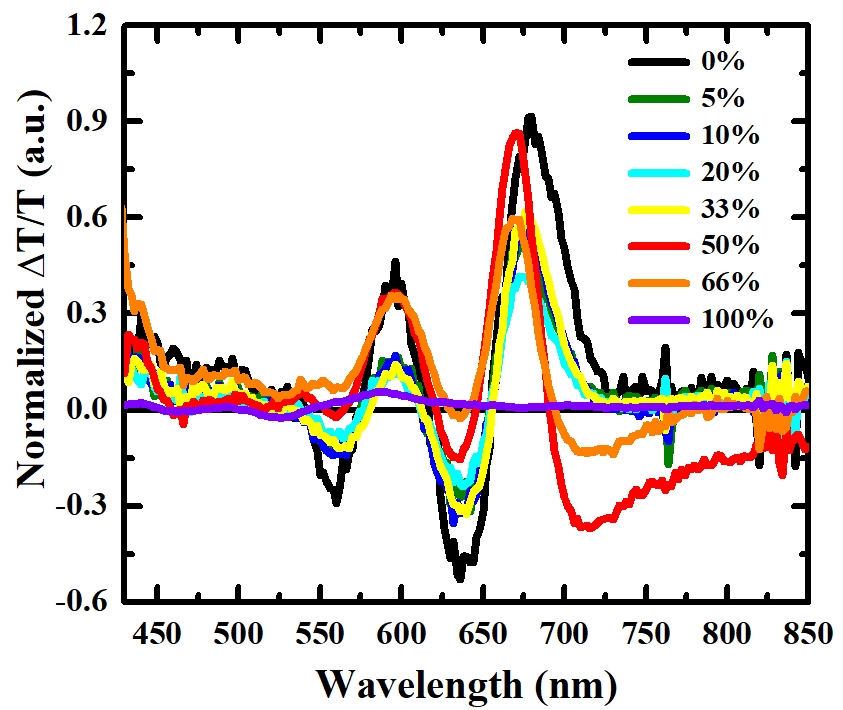}
	\caption{ The EA spectrum of pristine PPDT2FBT, PCBM and blends at 0 V DC bias. The spectra are normalized per unit applied field for comparison as the thickness of the different blends has slight variation.}
	\label{Fig:PPDT2FBT_Blend_EA_Normalized_1w}
\end{figure}

\begin{figure}[h!]
	\centering
	\includegraphics[width =0.85 \linewidth]{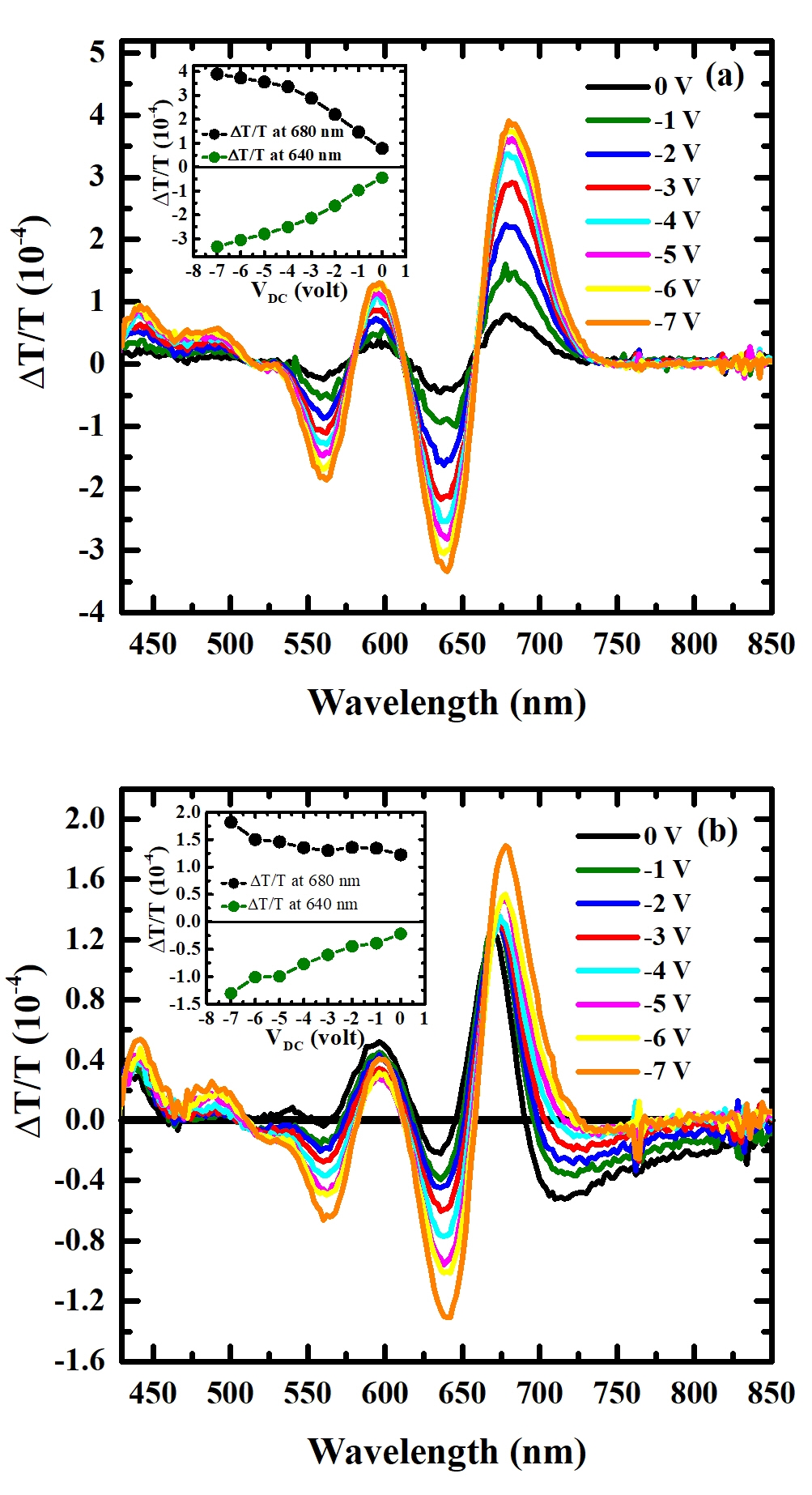}
	\caption{ The EA spectrum of (a) pristine PPDT2FBT and (b) blend having 50\% PCBM for different DC bias. The AC bias was 1.0 V (pp) and kept constant for all the spectra. The spectra were measured at first harmonic of applied AC bias frequency (1.3 kHz). The inset figures show the variation of EA signal with DC bias (reverse) at 680 nm (black) and 640 nm (olive green). }
	\label{Fig:PPDT2FBT_Blend_EA_pristine_and_50_50_1w}
\end{figure}

We carried out EA measurements on PPDT2FBT: PCBM blends with different compositions at room temperature in the transmission mode. The EA signals measured at the first harmonic of the applied AC bias frequency is expressed as,
\begin{equation}
\frac{\Delta T}{T} \propto \left(  F_{DC} - F_{BI} \right). F_{AC}
\end{equation}

where $\Delta T/T$ is the relative change in transmittance in the presence of applied bias, $F_{BI}$ is the built-in field of the device, and $F_{DC}$ and $F_{AC}$ are the applied DC and AC electric fields, respectively \cite{sahoo2021investigation}.

The EA spectra of different blend compositions, measured at first harmonic of the modulation frequency for 0 V DC bias, is shown in Figure \ref{Fig:PPDT2FBT_Blend_EA_Normalized_1w}. For the wavelength region below 650 nm, the spectral features were quite similar, for the different samples. The peak around 680 nm, however, showed an apparent blue shift and narrowing for a high PCBM percentage. These changes could be due to the increase in the negative dip above 720 nm, which is attributed to polaronic effects. Typically, polaron absorption is observed in the EA spectrum in the forward bias at lower energy than the HOMO-LUMO gap of the organic semiconductor \cite{zozoulenko2018polarons}. However, at zero DC bias, where the built-in voltage plays the role, significant polaron absorption can arise due to the trapping of photogenerated carriers in the bulk heterojunctions within the active layer and at the domain boundary.

\begin{figure*}
	\centering
	\includegraphics[width =0.7 \linewidth]{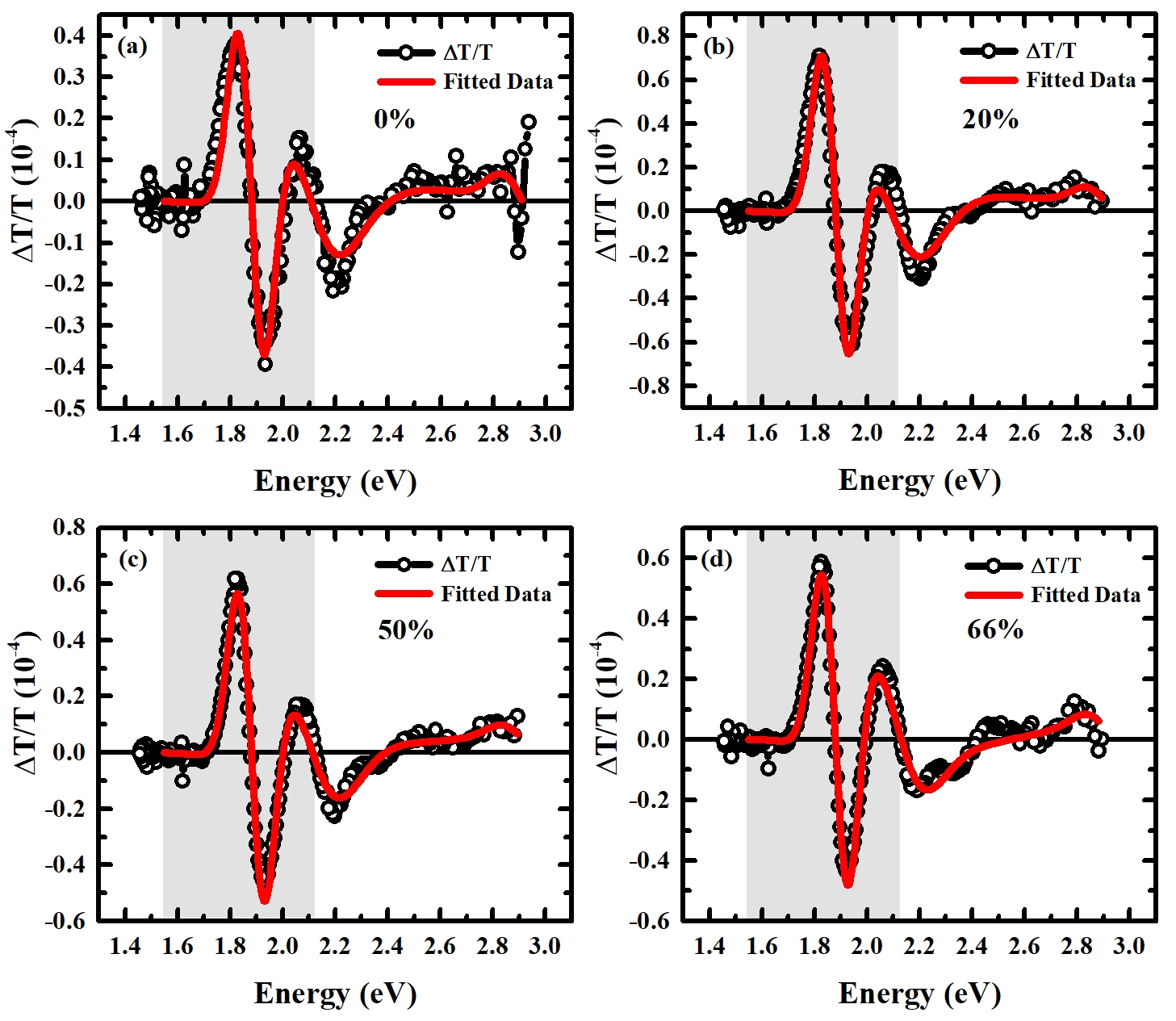}
	\caption{ The EA spectrum of (a) pristine PPDT2FBT, and the blends having (b) 20\%, (c) 50\% and (d) 66\% PCBM at the second harmonic of the modulation frequency. The applied AC and DC bias were 3 V (pp) and 0 V, respectively. The circles with solid line are the measured data, and the fitted data are shown by the solid red line. In the shaded region, the absorption contribution of PCBM is not significant. Therefore, the values of $\Delta P$ and $\Delta \mu$ corresponding to the absorption band in that region is considered for analysis.}
	\label{Fig:PPDT2FBT_Blend_EA_2w}
\end{figure*}

To study the effect of the bias voltage, EA spectra for different V$_{DC}$ were measured for thin film of pristine PPDT2FBT and blend having 50\% PCBM (Figure \ref{Fig:PPDT2FBT_Blend_EA_pristine_and_50_50_1w}). For the later, the polaronic feature decreased with higher reverse bias, and the blue shift and narrowing of the 680 nm peak vanished. This could be attributed to the depletion of the space charges, giving rise to polarons, due to the high bias \cite{sahoo2021investigation}. The change in the EA signal at 680 nm, however, was very small and not linear (Figure \ref{Fig:PPDT2FBT_Blend_EA_pristine_and_50_50_1w}(b) inset). For pristine PPDT2FBT films, the polaronic feature was absent in the EA spectrum, and variation of the DC bias resulted in the change of magnitude of the EA signal, with no change in lineshape. This was evident from the linear increase in the EA signal with $V_{DC}$, up to -4 V (Figure \ref{Fig:PPDT2FBT_Blend_EA_pristine_and_50_50_1w}(a) inset). For higher bias values, space charges influenced the EA signal, and deviation from the linear behaviour was observed.

These observations indicated that the EA line shapes for blended films had considerable influence from the polaronic effects. Extraction of parameter and material properties from EA measurements require fitting of the EA spectra. However, the lineshape distortion from standard excitonic contributions due to presence of polarons in the first harmonic EA signal, made it difficult to analyze. Such effects are absent in the second harmonic EA signal, which is proportional to the square of the applied AC field, given by,  

\begin{equation}
\left( \frac{\Delta T}{T} \right)_{2\omega} \propto \left( - \frac{1}{2} F_{AC}^{2} \right)
\label{equn:delta_T_by_T_2w_1}
\end{equation}

The equation \ref{equn:delta_T_by_T_2w_1} in terms of the derivative of the absorbance ($A(E)$) can be written as \cite{sahoo2022investigation} 
\begin{equation}
\left( \frac{\Delta T}{T} \right)_{2\omega}  = 2.303 \times \frac{\pi}{2} \times [rA(E) + pA^{'}(E)  + qA^{''}(E)]
\end{equation}

where r, p, and q are coefficients of the zero field absorbance and its first and second order derivative, respectively, contributing to the EA spectrum. The coefficients are related to the anisotropic distribution of transition dipole moments in the material, the change in polarizability ($\Delta P$), and change in dipole moment ($\Delta \mu$) in the system, as discussed details in ref \cite{jalviste2007theoretical}. The coefficient $p$ and $q$ can be expressed by equation \ref{equn:expression_p_2w} and \ref{equn:expression_q_2w} when measured at the second harmonic of the applied modulation frequency.
\begin{equation}
p = \frac{1}{2} \Delta P |\frac{1}{2} F_{AC}^{2} |
\label{equn:expression_p_2w}
\end{equation}

\begin{equation}
q = \frac{1}{6} (\Delta \mu )^2 |\frac{1}{2} F_{AC}^{2} |
\label{equn:expression_q_2w}
\end{equation}

Thus, fitting of the EA spectrum gives values of $r$, $p$, and $q$, which will help estimate the corresponding $\Delta P$ and $\Delta \mu$, using the above equations. The magnitudes of $\Delta P$ and $\Delta \mu$  are indicative of the nature of excitons contributing to certain features in the EA spectrum.

\begin{figure}[h!]
	\centering
	\includegraphics[width =0.85\linewidth]{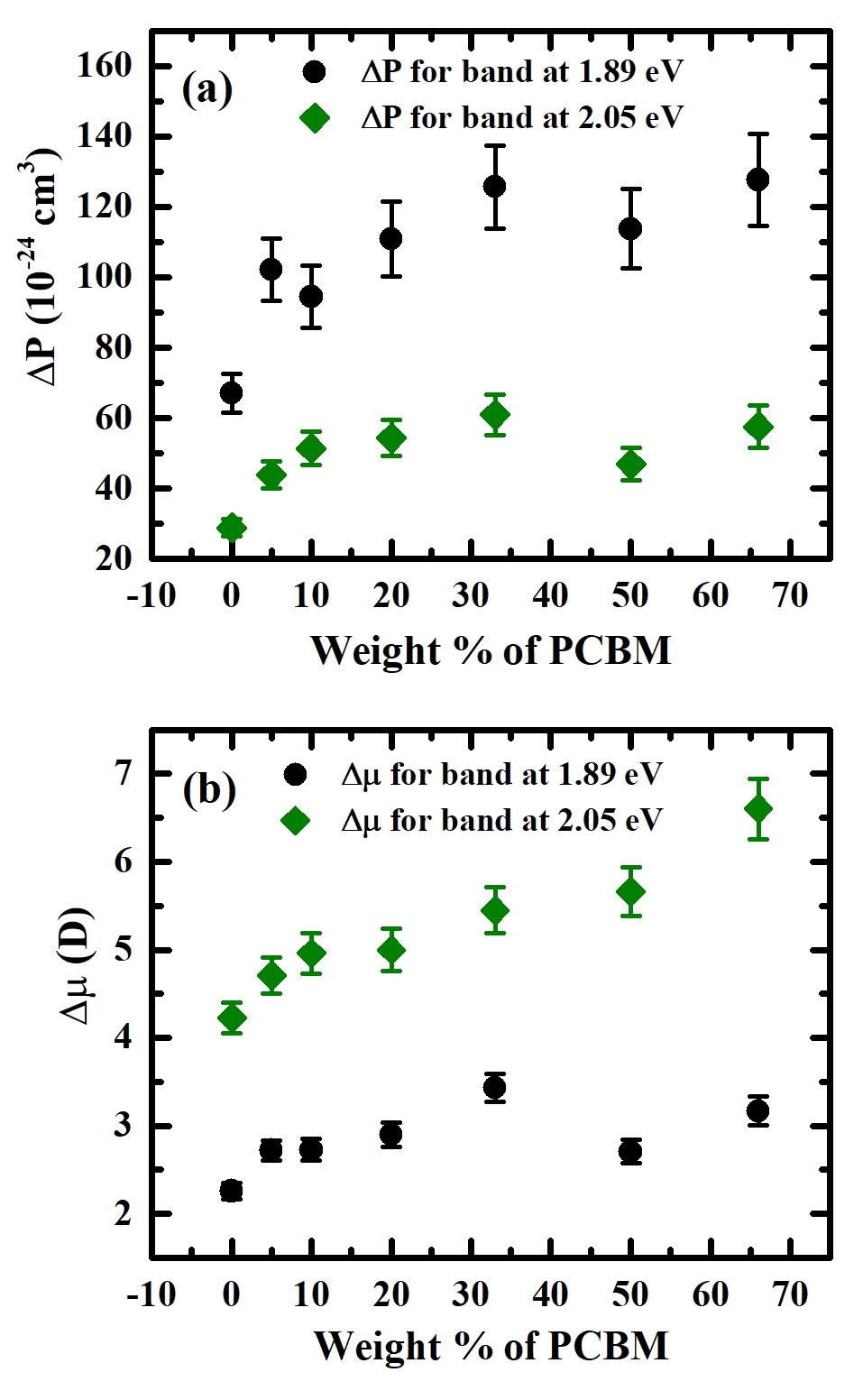}
	\caption{ The value of (a) $\Delta P$ and (b) $\Delta \mu$ of pristine PPDT2FBT and the blends associated with first two absorption bands. The error bar of $\Delta P$ and $\Delta \mu$ arise primarily due to the error in thickness of the film of each blend composition.  } 
	\label{Fig:PPDT2FBT_Blend_delta-p_delta_mu}
\end{figure}

The EA spectrum of pristine PPDT2FBT and different blends were measured at second harmonic of the modulation frequency. In Figure \ref{Fig:PPDT2FBT_Blend_EA_2w}, the EA spectrum of pristine PPDT2FBT and three different blends (having 20\%, 50\% and 66\% PCBM) are shown. The shape of the EA spectrum of the blends and pristine PPDT2FBT were quite similar, suggesting that the contribution of PCBM absorption in the EA spectrum of the blends was negligible. The effect of PCBM in the blend was primarily in modifying the nature of excitons of PPDT2FBT by enabling the efficient formation of CT excitons. We, therefore, used the absorption spectrum of pristine PPDT2FBT film of thickness $d$, and its derivatives to fit the measured EA spectra to determine the values of $\Delta P$ and $\Delta \mu$. The absorption spectrum of the PPDT2FBT was reconstructed using six Gaussian bands (Supplementary Figure S1). We restricted our analysis between 800 nm and 585 nm i.e. first two absorption bands (shaded regions in Figure \ref{Fig:PPDT2FBT_Blend_EA_2w}). The fitted EA spectra are shown in Figure \ref{Fig:PPDT2FBT_Blend_EA_2w}. The obtained values of  $\Delta P$ and $\Delta \mu$ for different blend compositions are plotted in Figure \ref{Fig:PPDT2FBT_Blend_delta-p_delta_mu}.   

\begin{figure}[h!]
	\centering
	\includegraphics[width = 0.9\linewidth]{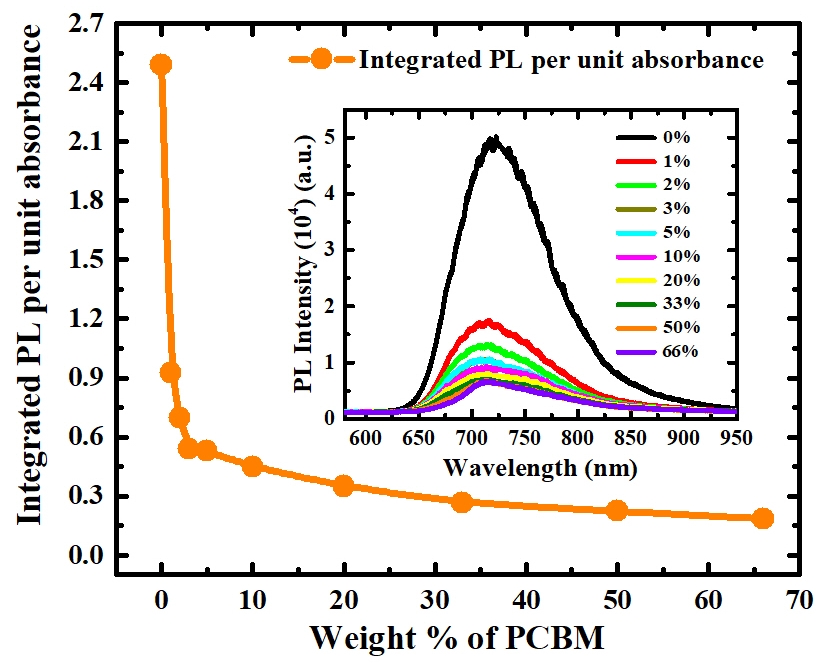}
	\caption{ The integrated PL as a function of weight percent of PCBM. The integrated PL was corrected for difference in absorption of different blends. The inset figure shows the PL spectrum of pristine PPDT2FBT film and blends. }
	\label{Fig:PPDT2FBT_Blend_PL}
\end{figure}

For both the bands at 657 nm (1.89 eV) and 606 nm (2.05 eV), $\Delta P$ and $\Delta \mu$ are increased in blends compared to the pristine PPDT2FBT for both the bands. This increase in $\Delta \mu$ with PCBM indicates the CT excitons formed at the PPDT2FBT/PCBM interface become more delocalized and have lower binding energy. Therefore, the dissociation probability of the excitons increases in presence of PCBM. Hence, it is expected that the radiative recombination of the excitons would be suppressed in the presence of PCBM. The integrated photoluminescence of the blend having 1\% PCBM reduced drastically ($\sim$ 63\%) compared to pristine PPDT2FBT (Figure \ref{Fig:PPDT2FBT_Blend_PL}). Eventually, the PL quenching reached saturation for PCBM concentrations at about 20\%. 

\begin{figure}[h!]
	\centering
	\includegraphics[width = 0.8\linewidth]{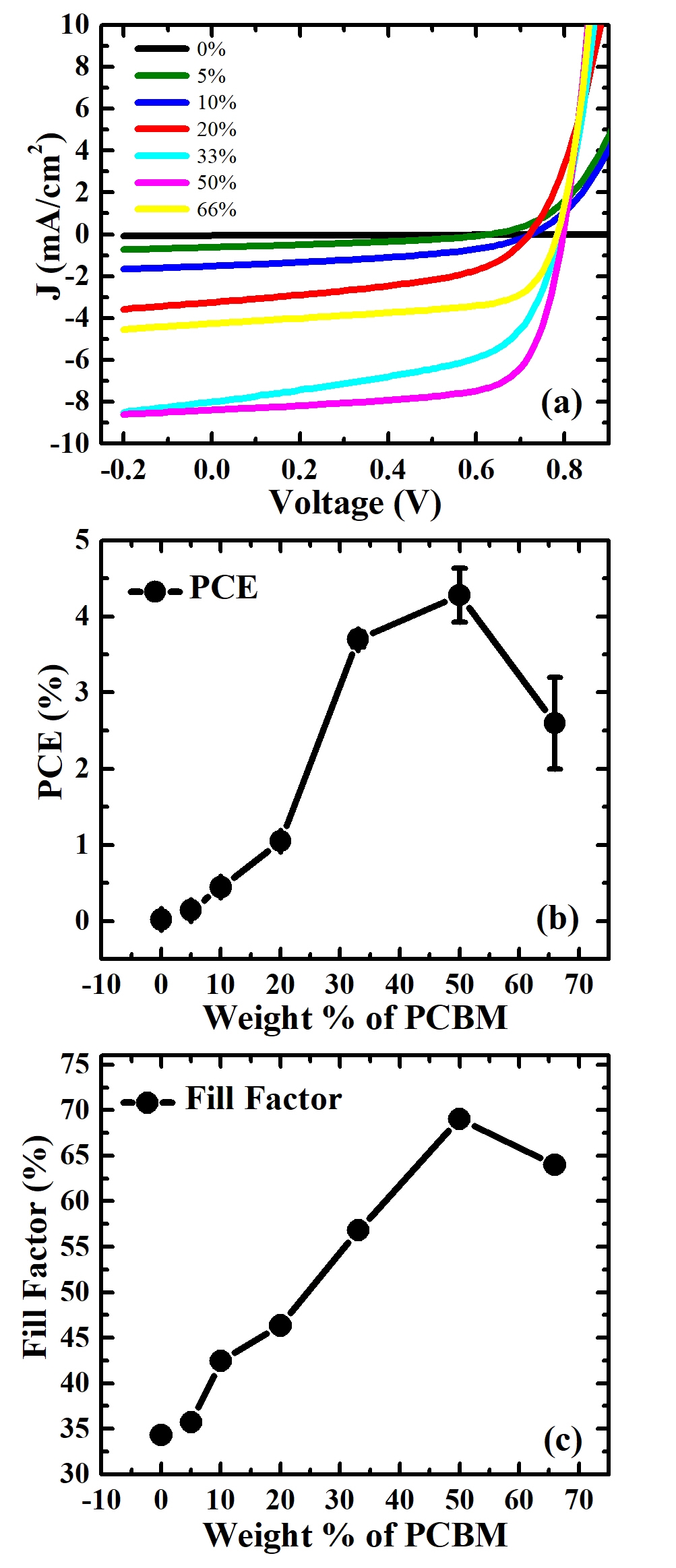}
	\caption{ (a) The photocurrent of the devices having different percentage of PCBM under 1 sun illumination. (b) The average PCE of the devices for different percentage of PCBM are plotted. The error bar shows the variation of PCE over multiple measurements in four devices for each blend composition. (c) The fill factor of the devices for different percentage of PCBM. }
	\label{Fig:PPDT2FBT_Solar_Cell_J_SC_FF_PCE}
\end{figure}

Since the PL quenching can be an indication of the generation of free carriers in the system, improvement in photocurrent generation efficiency is expected in the presence of PCBM. We studied the photocurrent efficiency of devices with active layers consisting of blends of PPDT2FBT: PCBM with different compositions. Measurements were done for devices (ITO/PEDOT: PSS (30 nm) / PPDT2FBT: PCBM/ LiF (1 nm) / Al (80 nm)) under 1 sun illumination (AM1.5). The photocurrent of the devices improved with increasing PCBM content in the active layer (Figure \ref{Fig:PPDT2FBT_Solar_Cell_J_SC_FF_PCE}(a)). Maximum short circuit current was observed for the devices having 33\% and 50\% PCBM. However, the power conversion efficiency (PCE) (Figure \ref{Fig:PPDT2FBT_Solar_Cell_J_SC_FF_PCE}(b)) of the device having 50\% PCBM was much higher than that of the 33\% PCBM device. This was attributed to the higher fill factor of the 50\% PCBM device, as shown in Figure \ref{Fig:PPDT2FBT_Solar_Cell_J_SC_FF_PCE}(c).

The generated photocurrent in the pristine PPDT2FBT device was poor compared to other devices. The addition of 5\% PCBM with PPDT2FBT improved the photocurrent by one order. This increase in photocurrent was the signature of the better dissociation of the generated excitons in the presence of PCBM, which was manifested as PL quenching. Although a substantial quenching (79\%) in PL was observed for the device having 5\% PCBM, the PCE of the device was not significantly increased. The PCE increased monotonically till the PCBM content was as high as 50\%, though PL quenching had reached saturation at about 20\% PCBM (Figure \ref{Fig:PPDT2FBT_Blend_PL}). This indicated that factors other than exciton dissociation play a significant role in improving photocurrent generation efficiency. One such factor is the collection efficiency of the photogenerated carries \cite{amorim2020analytical}.

\begin{figure*}
	\centering
	\includegraphics[width = \linewidth]{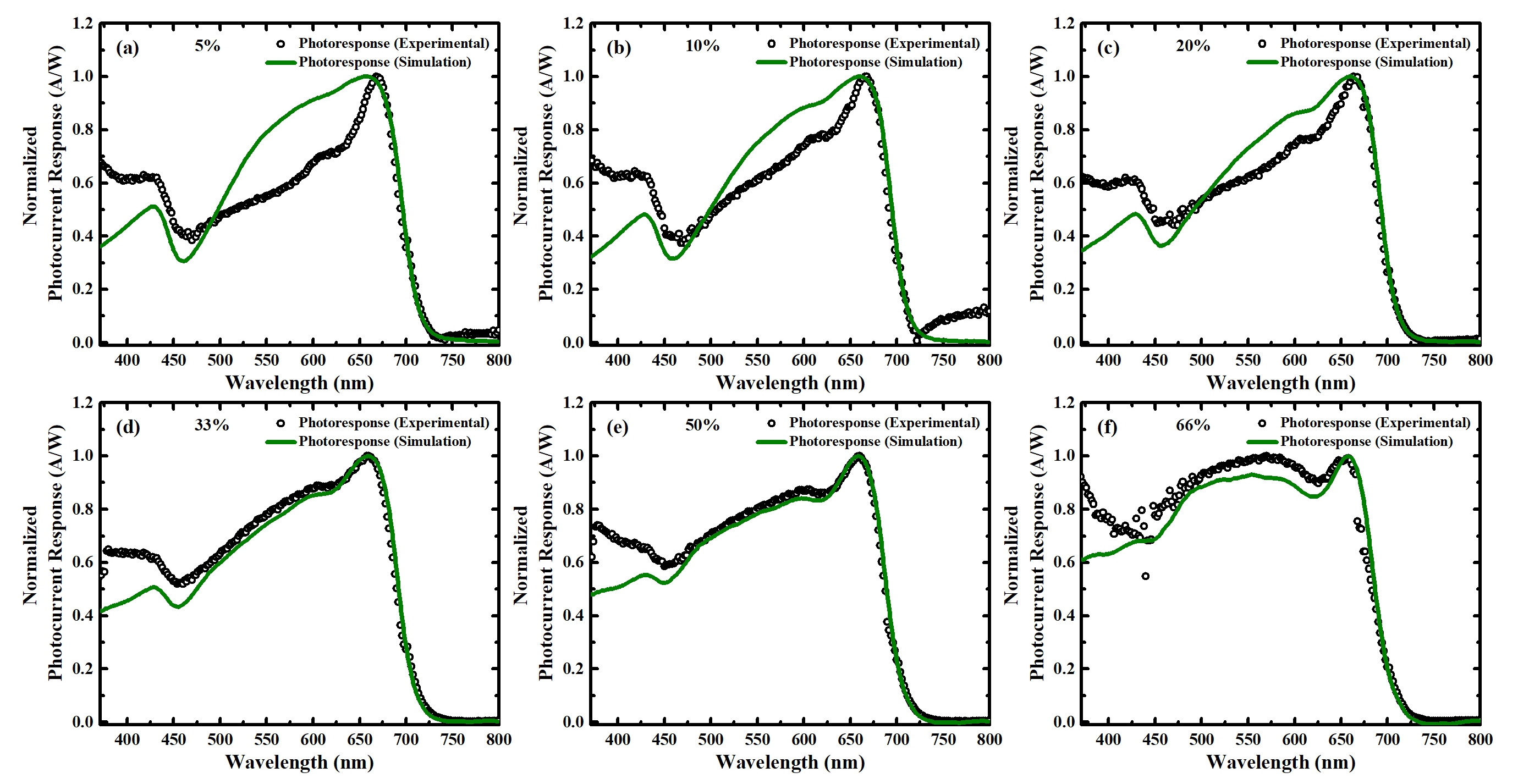}
	\caption{ The normalized photocurrent response spectra of the devices having (a) 5\%, (b) 10\%, (c) 20\%, (d) 33\%, (e) 50\% and (f) 66\% PCBM. The measured and calculated spectra are shown by open circles and solid olive green line respectively.}
	\label{Fig:PPDT2FBT_Blend_PC}
\end{figure*}

The collection efficiency depends largely on the hole, and electron mobility in the active layer \cite{ray2011quantitative}. The morphology of the film also plays an important role in the collection of the carriers \cite{szarko2014photovoltaic}. Similar values of electron and hole mobilities are desirable for better transport and collection of the carriers from the bulk of the active layer. For materials having a large difference between electron and hole mobilities, the interfacial carriers primarily contribute to the photocurrent if the collection length is less than the thickness of the active layer \cite{ray2011quantitative}. In PPDT2FBT the mobility of holes are high \cite{nguyen2014semi}. However, the addition of PCBM improves the electron mobility in the blend due to the high intrinsic electron mobility of PCBM \cite{ebenhoch2015charge}. The morphology of the film may also change with the amount of PCBM in the blend.  And the number of PPDT2FBT and PCBM interfaces also increases with the PCBM content. These interfaces not only help in the dissociation of the excitons but also improves the transport of the electrons through the bulk of the film. In pristine PPDT2FBT and the blends having less amount of PCBM, a major part of the dissociated carriers may be lost by non-radiative processes or captured by the traps before being collected by the electrodes leading to low PCE.

The carrier collection efficiency is reflected in the fill factor of the device. The calculated fill factors of the devices from the experimental data are shown in Figure \ref{Fig:PPDT2FBT_Solar_Cell_J_SC_FF_PCE}(c). It increased with the increasing percentage of PCBM, suggesting that the PCBM in the blends helped in the transport of the carriers. The lower fill factor and PCE for the device having 66\% PCBM compared to those of the 50\% PCBM device might be due to the reduction of effective absorption coefficient and the number of PPDT2FBT-PCBM interfaces for such high PCBM content. 

The photocurrent extracted from the device can originate from the excitons generated in the bulk of the active layer or at the electrode/film interfaces \cite{ray2006bulk,ray2011quantitative, ghosh1978merocyanine}. The exciton generation is proportional to the absorption of the material. The absorption of incident light of various wavelengths at a certain depth of the active layer is different. Therefore, the absorption spectrum at various depths of the device would contribute differently to the total absorption spectrum of the active layer. Hence, the shape of the total absorption spectrum is different from the absorption spectrum from a certain depth. Since, the photocurrent spectrum is proportional to $\lambda \mathcal{A}(\lambda)$ assuming 100\% internal quantum efficiency at all wavelength, where $\mathcal{A}(\lambda)$ is the absorptance of the device for light of wavelength $\lambda$ \cite{bhat2020high}, the shape of the photocurrent spectrum varies depending on the origin of the dissociated carriers \cite{ray2011quantitative}. 

For bulk generation, the absorptance has contributions from the entire thickness ($d$) of the film, and is given by  $\mathcal{A}(\lambda)_{bulk} = \mathcal{A} (\lambda, 0 \leq x \leq d) = \int_{0}^{d} \mathcal{A}(\lambda, x) dx $. For photocurrent generation from exciton dissociation at the interface, layers close to the interface at depth $L$, contribute (where $L<d$). Thus, the corresponding absorptance can be written as $ \mathcal{A}(\lambda, L \leq x \leq d) = \int_{L}^{d} \mathcal{A}(\lambda, x) dx $. In our simulations of photocurrent, $\mathcal{A}(\lambda)$ was calculated using transfer matrix method \cite{ohta1990matrix} where bulk generation was considered. Interference effects were taken into consideration since this effect was considerable in the device structures due to the high reflectivity of the aluminium electrodes. The optical constants ($n$ and $k$) of the thin films, used for the calculations, are shown in the supplementary section S2. Comparison of simulations with experimental measurements (Figure \ref{Fig:PPDT2FBT_Blend_PC}) showed that for devices with low PCBM contents, there was a considerable mismatch between calculated and measured photocurrent spectra, primarily between 500 and 650 nm. This implied that for low PCBM concentrations bulk generation model did not hold, i.e. then the carriers generated in the bulk of the active layer were not collected effectively to contribute to the photocurrent. The interfacial charges could be responsible for the photocurrents in these devices. For other devices with higher PCBM concentrations (Figures \ref{Fig:PPDT2FBT_Blend_PC} d, e, f), the measured and calculated spectra had similar line shapes, indicating that the photocurrents in these devices were dominated by the bulk photogenerated carriers. Thus, an increase in PCBM percentage in the blends, facilitated better transport of carriers through the bulk. This gave a comprehensive picture of the role played by PCBM in the blended system. PCBM helped in the dissociation of the excitons and also increased the collection efficiency of these carriers throughout the bulk of the active layer.

\section{\label{sec:level4}Conclusion}

In this work, we measured the EA spectrum of PPDT2FBT: PCBM blend and determined the strength of Frenkel and CT characteristics in blend compared to pristine system. No new CT states were observed in the blends below the band edge energy. However in presence of PCBM, the polarizability and dipole moment of the excitons near the band edge increased, indicated higher exciton dissociation probability, which manifested as drastic PL quenching in PL measurements in the blends. The blend having 5\% PCBM showed $\sim$ 81\% PL quenching compared to pristine PPDT2FBT.  However, the photocurrent efficiency increased with further PCBM content in the blend, and the highest PCE was obtained for 50\% PCBM. This increase was attributed to better transport of carriers in the active layer. Simulations proved that, for a higher percentage of PCBM, carriers from the bulk of the film contributed to the photocurrent. This study corroborated that PCBM played a dual role in PPDT2FBT: PCBM blended solar cell device - (i) helped in exciton dissociation and (ii) facilitated transport of the generated carrier through the whole thickness of the film.

\section*{\label{sec:level6}Data availability statement}

The data that support the findings of this study are available from the corresponding author upon reasonable request.

\section*{\label{sec:level5}Acknowledgments}
The authors acknowledge the Center for NEMS and Nanophotonics (CNNP) at IIT Madras for facilitating the device fabrication. The authors would like to acknowledge Council of Scientific \& Industrial Research, India, for financial support (No. 03(1465)/19/EMR-II).

\section*{References}
\bibliographystyle{unsrt}
\bibliography{aip_Ref.bib}

\end{document}